\documentclass{rmf-d}
\usepackage{nopageno,rmfbib,multicol,times,epsf,amsmath,amssymb,cite}
\usepackage[latin1]{inputenc}
\usepackage[]{caption2}
\usepackage{graphics}
\usepackage{slashed}
\usepackage[latin1]{inputenc}
\usepackage[demo]{graphicx}
\usepackage{hyperref}
\usepackage{lipsum}
\usepackage{wrapfig, blindtext}
\usepackage{comment}
\usepackage{tabularx}
\usepackage{float}
\usepackage{tikz}
\usepackage{verbatim}
\usetikzlibrary{mindmap,arrows,shapes,trees}
\usetikzlibrary{shapes.multipart}
\usepackage[compat=1.1.0]{tikz-feynman}
\usepackage{contour}
\usepackage{feynmp}
\usepackage{tikzpagenodes}
\usepackage{float}
\tikzstyle{overall} = [rectangle, rounded corners, anchor=text, text centered, draw=black, rectangle split, rectangle split parts=2]
\tikzstyle{start} = [rectangle, rounded corners, minimum width=3cm, minimum height=1cm,text centered, draw=black, fill=blue!10]
\tikzstyle{DD} = [rectangle, rounded corners, minimum width=3cm, minimum height=1cm,text centered, draw=black, fill=red!10]
\tikzstyle{process} = [rectangle, rounded corners, minimum width=3cm, minimum height=1cm, text centered, draw=black, fill=orange!20]
\tikzstyle{inbox} = [rectangle, minimum width=0.75cm, minimum height=0.35cm, text centered, draw=black, fill=pink!20]
\tikzstyle{stop} = [rectangle, rounded corners, minimum width=3cm, minimum height=1cm,text centered, draw=black, fill=gray!10]
\tikzstyle{arrow} = [thick,->,>=stealth,line width=2pt]
\tikzstyle{DDRT} = [rectangle, rounded corners, minimum width=2cm, minimum height=0.5cm,text centered, draw=black, fill=gray!10]
\tikzstyle{startRT} = [rectangle, rounded corners, minimum width=2cm, minimum height=0.5cm, text centered, draw=black, fill=red!10]
\tikzstyle{processRT} = [rectangle, minimum width=1cm, minimum height=0.5cm, text centered, draw=black, fill=orange!10]
\tikzstyle{stopRT} = [rectangle, rounded corners, minimum width=2cm, minimum height=0.5cm,text centered, draw=black, fill=blue!10]
\tikzstyle{DashBox} = [rectangle, rounded corners, minimum width=3.5cm, minimum height=2.75cm, draw=black, dashed]
\def\rmfcornisa{Research OR Education of Physics \hfill\rmf\hfill December 2021}

\def\rmfcintilla{{\it Rev.\ Mex.\ Fis.\/}}
\clearpage \rmfcaptionstyle 
\begin{document}
\title{ Accessing pion GPDs through the Sullivan process: is it feasible? 
\vspace{-6pt}}
\author{J. M. Morgado}
\address{Dpto. Ciencias Integradas and CEAFMC, Universidad de Huelva, E-21071 Huelva, Spain}
\author{V. Bertone}
\address{Irfu, CEA, Universit\'e Paris-Saclay, 91191, Gif-sur-Yvette, France}
\author{M. Defurne}
\address{Irfu, CEA, Universit\'e Paris-Saclay, 91191, Gif-sur-Yvette, France}
\author{F. De Soto}
\address{Dpto. Sistemas F\'isicos, Qu\'imicos y Naturales, Universidad Pablo de Olavide, E-41013 Sevilla, Spain}
\author{C. Mezrag}
\address{Irfu, CEA, Universit\'e Paris-Saclay, 91191, Gif-sur-Yvette, France}
\author{H. Moutarde}
\address{Irfu, CEA, Universit\'e Paris-Saclay, 91191, Gif-sur-Yvette, France}
\author{J. Rodr\'iguez-Quintero}
\address{Dpto. Ciencias Integradas and CEAFMC, Universidad de Huelva, E-21071 Huelva, Spain}
\author{J. Segovia}
\address{Dpto. Sistemas F\'isicos, Qu\'imicos y Naturales, Universidad Pablo de Olavide, E-41013 Sevilla, Spain}
\maketitle
\begin{abstract}
\vspace{1em} Describing hadronic structure is one of the most intriguing problems in physics. In this respect, generalized parton distributions (GPDs) constitute an outstanding tool, allowing to draw ``three dimensional pictures" of hadron's inside. Starting from contemporary models for pion's GPDs fulfilling all constraints imposed by QCD, we compute Compton form factors of pions subjected to deeply virtual Compton scattering. We show the behaviour of Compton Form Factors (CFFs) to be gluon-dominated at EIC's kinematics. Finally we evaluate lepton-beam-spin asymmetries in the Sullivan process, demonstrating the existence of such and thus triggering optimism about the possibility of probing pion's $3$D structure at electron-ion colliders. \vspace{1em}
\end{abstract}
\keys{Non-perturbative QCD, DVCS, CFFs, EIC, Sullivan process. \vspace{-4pt}}
\pacs{\bf{\textit{PLEASE PROVIDE}}    \vspace{-4pt}}
\begin{multicols}{2}

\section{Introduction}


Among all hadrons, pions are one of the most intriguing systems of study  \cite{Aguilar:2019teb, Roberts:2021nhw}. As Nambu-Goldstone bosons of QCD's dynamical breakdown of chiral symmetry, they constitute a unique laboratory for the study of emergent strong phenomena. Moreover, as ``two-bodies" bound-states they are the simplest systems tied by the strong interaction. Therefore, gaining insights into pion's structure is of great interest \cite{Aguilar:2019teb, Montgomery:2017yua, Arrington:2021biu, AbdulKhalek:2021gbh, Anderle:2021wcy, Adams:2018pwt} and the most direct tool we have for this purpose is lepton scattering.

Unfortunately targeting pions is a really challenging task, requiring indirect approaches to be explored in order to probe pion's structure. In this respect, the so-called Sullivan process \cite{Sullivan:1971kd} provides a remarkable tool. In a nutshell, given that the necessary conditions are met (see Sec. \ref{sec:SullivanProcess}), the paradigmatic case of electron-proton scattering can also take place through Deeply Virtual Compton Scattering (DVCS) with a pion in the meson cloud of the nucleon; opening a clear window to the study of pion's structure. Moreover two mechanisms contribute to such process \cite{Amrath:2008vx}: Bethe-Heitler (BH) and DVCS, which is related with Generalized Parton Distributions (GPDs) \cite{Ji:1996nm}.

GPDs were introduced more than two decades ago as extensions for the ``classical" Parton Distribution Functions (PDFs) to off-forward kinematics \cite{Mueller:1998fv, Ji:1996nm, Radyushkin:1997ki, Diehl:2003ny}. They were shown to offer access to the three-dimensional structure of hadrons \cite{Burkardt:2000za} and a direct connection with QCD's fundamental degrees of freedom through the energy-momentum tensor \cite{Ji:1996ek}.

As interesting as they are, GPDs are notoriously difficult to extract from experiment and theory \cite{Bertone:2021yyz, Bertone:2021wib}. The first models for pion GPDs fulfilling all the requirements from quantum field theory are contemporary \cite{Chouika:2017dhe, Chavez:2021llq}, but the possibility of accessing them at future experimental facilities demands further scrutiny. We herein report on a recent effort addressing this last point \cite{Chavez:2021koz}.

\section{Accessing pion GPDs: the Sullivan process}\label{sec:SullivanProcess}

As shown in a seminal paper by J. D. Sullivan \cite{Sullivan:1971kd}, the cross-section for exclusive electron-nucleon scattering in the Bjorken limit receives sizeable contributions from pion-nucleon final states when the momentum transferred between the initial and final nucleon states, $t$, remains small and near the threshold for pion production: $\left|t\right|\lesssim 0.6~\text{GeV}^{2}$ \cite{Qin:2017lcd}. Under these conditions, such process can be interpreted in the so-called \textit{one-pion-exchange} approximation, \textit{i.e.} in terms of pion-photon scattering amplitudes. Moreover, the invariant mass of the nucleon-$\pi$ system being cut as $M_{N\pi}^{2}\gtrsim 4~\text{GeV}^{2}$, possible contamination from nucleon resonances is minimized \cite{Sullivan:1971kd, Amrath:2008vx}, allowing to interpret the Sullivan process in terms of direct interaction between a photon and a slightly off-shell pion.

Following these arguments, the amplitude for the Sullivan process ($ep\rightarrow e\pi n\gamma$) can be ``factorized" as:
\begin{equation}
\mathcal{M}_{ep\rightarrow e\pi n\gamma}=\mathcal{M}_{p\rightarrow\pi n}\mathcal{M}_{e\pi\rightarrow e\pi\gamma}
\end{equation}
drawing a clear path toward the experimental analysis of pion's structure through the amplitude for $e\pi\rightarrow e\pi\gamma$.

Two subprocesses give contributions to such amplitude \cite{Amrath:2008vx}: Bethe-Heitler scattering, which provides direct access to the pion's Electromagnetic Form Factor (EFF) and has already been exploited for its extraction at large photon virtualities \cite{Huber:2008id}. And a DVCS contribution:
\begin{equation}\label{eq:Contributions}
\mathcal{M}_{e\pi\rightarrow e\pi\gamma}=\mathcal{M}_{\text{BH}}+\mathcal{M}_{\text{DVCS}}
\end{equation}

DVCS amplitudes are parametrized by Compton Form Factors (CFFs) \cite{Ji:1996nm}. These being defined as convolutions of a perturbatively calculable kernel with GPDs \cite{Belitsky:2010jw, Pire:2011st, Radyushkin:1997ki, Collins:1998be, Ji:1998xh}, the DVCS contribution to $e\pi\rightarrow e\pi\gamma$ allows for a direct study of pion's three-dimensional structure.

Cross sections being defined as the modulus squared of the corresponding amplitude, here Eq. \eqref{eq:Contributions}, three contributions to that of the Sullivan process are identified: pure DVCS (BH), parametrized by modulus squared CFFs (EFFs); and a third contribution, the interference term between two such subprocesses, directly related with the real and imaginary parts of the CFFs \cite{Amrath:2008vx, Belitsky:2010jw}. Moreover, at leading twist, pure DVCS and BH contributions exhibit no dependence on lepton-beam-spin polarization while the interference term does \cite{Amrath:2008vx}. Therefore, the lepton's Beam Spin Asymmetry (BSA), $\mathcal{A}$:
\begin{equation}\label{eq:LBSA}
\mathcal{A}:=\left.\frac{\sigma^{\uparrow}-\sigma^{\downarrow}}{\sigma^{\uparrow}+\sigma^{\downarrow}}\right|_{e\pi\rightarrow e\pi\gamma}\hspace{-1mm}=\frac{\mathcal{S}_{\text{Im}}\text{Im}\left(\mathcal{H}_{\pi}\right)}{\text{BH}+\text{DVCS}+\mathcal{C}_{\text{Re}}\text{Re}\left(\mathcal{H}_{\pi}\right)}
\end{equation}
grants access to the real and imaginary parts of CFFs, $\mathcal{H}_{\pi}$; thus allowing to employ the Sullivan process to measure them (at least formaly) or, equivalently, to gain insights into the pion's three-dimensional structure through GPDs.

Furthermore, the coefficients of each term in Eq. \eqref{eq:LBSA} are shown to be proportional to $\cos\phi_{\pi}$ ($\mathcal{C}_{\text{Re}}$) and $\sin\phi_{\pi}$ ($\mathcal{S}_{\text{Re}}$) \cite{Amrath:2008vx, Belitsky:2001ns}, with $\phi_{\pi}$ the angle between the planes defined by (incoming and outgoing) leptons and pions in the center of mass frame of $\pi\gamma$ in the final state \cite{Bacchetta:2004jz}. This means that not only ratios $\text{Im}\left(\mathcal{H}_{\pi}\right)/\text{Re}\left(\mathcal{H}_{\pi}\right)$ can be measured through the Sullivan process, but also to exploit the $\phi_{\pi}$ dependence of its interference to separately access the real and imaginary parts of CFFs.

The clarity of the above described procedure does not preclude the feasibility of a practical analysis of pion's structure to be obscured by experimental artefacts. Given that various experimental facilities devoted to the study of hadron structure are planned \cite{AbdulKhalek:2021gbh, 	Anderle:2021wcy, Adams:2018pwt}, a feasibility study of the methodology presented herein is timely.

Thus we choose to perform an analysis of the Sullivan process at the future Electron-Ion Collider (EIC) \cite{AbdulKhalek:2021gbh}. Relying on state-of-the-art models for pion GPDs \cite{Chavez:2021llq}, we evaluate Compton form factors at Next-to-Leading Order (NLO) in the strong coupling constant. From their analysis we demonstrate that pion's response to Compton scattering in the Bjorken limit is gluon-dominated. Finally we compute lepton-beam-spin asymmetries, showing that a clearly non-zero asymmetry might be expected at the foreseen EIC, thus triggering optimism about the possibility of probing pion's three-dimensional structure \cite{Chavez:2021koz}.

\section{Modelling of pion GPDs}\label{sec:GPDmodel}

The first ingredient needed for the present study is a reliable model for pion's $3$D structure. However, modelling of hadron GPDs is a difficult task. Mainly because their dependence on: $x$, average momentum fraction of the active quark; $\xi$ (skewness), half the longitudinal momentum transferred; and $t_{\pi}$, squared momentum transfer between pion states, are constrained by a set of properties \cite{Diehl:2003ny, Belitsky:2005qn}. In fact, only modern strategies have been able to build models fulfilling all of them by construction \cite{Chouika:2017dhe, Chavez:2021llq}. It is precisely such approach to which we stick.

\begin{figure}[H]
\centering
\begin{tikzpicture}[node distance=1.4cm]
\node[name=hola, align=center] {
    	\tikz{
    		\node[draw, align=center](outbox) (start) [process] {\textbf{PDF}\\
			\small{Non-perturbative QCD}    		
     		};    
		    \node (med)   [draw, align=center, DD, below of=start] {\textbf{DGLAP GPD}\\
		    \small{Positivity}  
		    };
    		
		    \node (stop)  [start, below of=med, align=center] {\textbf{ERBL GPD}\\
			\small{Polynomiality}
		    };
		    
	    	\draw [arrow] (start) -- (med);
	    	\draw [arrow] (med) -- (stop);
		    
			\node (RTmed) [draw, DDRT] at (3.5,-2.1) {DD};		
			
			\node (intRT) [processRT] at (3.5, -1.5) {$RT^{-1}$};
			\node (startRT) [startRT] at (3.5, -0.9) {\textbf{DGLAP}};
		    
		    \node (int2) [processRT] at (3.5, -2.7) {$RT$};
		    \node (stopRT)  [stopRT] at (3.5, -3.3) {\textbf{ERBL}};
		    
		    \node (Box) [DashBox] at (0,-2.1) {};
		    
		    \draw [decorate,decoration={brace,amplitude=10pt},xshift=1.9cm,yshift=-4cm] (0.4,0.4) -- (0.4,3.35) node [black,midway,xshift=-0.6cm] {};
    	}

};
\end{tikzpicture}
\caption{\label{fig:ModellingStrategy} Diagram describing contemporary modelling strategies for pion GPDs allowing to fulfil every fundamental property required by quantum field theory as introduced in \cite{Chavez:2021llq}.}
\end{figure}
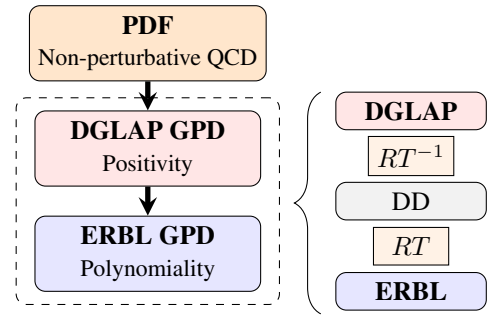

In summary, under the assumption of chiral symmetry, DGLAP GPDs ($\left|x\right|\geq\left|\xi\right|$) can be built from the usual PDFs as \cite{Zhang:2021mtn, Raya:2021zrz, Chavez:2021llq}: $H_{\pi}\left(x,\xi,t_{\pi}\right)=q_{\pi}\left(x\right)\Phi\left(z\right)$, where  $z=-t_{\pi}\left(1-x\right)^{2}/\left(1-\xi^{2}\right)$. Strikingly the GPDs thus defined preserve the positivity property \cite{Chavez:2021llq}. Then, the covariant extension formalism \cite{Chouika:2017dhe, Chouika:2017rzs} can be employed for the kinematic completion of the model within the ERBL domain ($\left|x\right|\leq\left|\xi\right|$); polynomiality and support properties being also satisfied (Fig. \ref{fig:ModellingStrategy}).

According to \cite{Chavez:2021llq}, modelling of pion GPDs is therefore reduced (in a first stage) to that of the PDF. In this respect we choose to start from the results presented in \cite{Ding:2019lwe}, where realistic \textit{Ans\"atze} for the pion's PDF were obtained from continuum Schwinger methods, showing a large-$x$ behaviour which remains compatible with that extracted from experimental data including soft-gluon (threshold) resummation \cite{Barry:2021osv, Aicher:2010cb, Cui:2021mom}. 

The corresponding DGLAP GPD model was presented in \cite{Raya:2021zrz}, and further elaborated with its kinematic completion in \cite{Chavez:2021llq}. It was also shown to fulfil by construction with the support property of GPDs \cite{Diehl:1998sm};  with the bounds imposed by positivity of the underlying Hilbert-space norm \cite{Pire:1998nw, Radyushkin:1998es, Diehl:2000xz, Pobylitsa:2002iu}, and also to preserve Lorentz invariance, as exhibited through polynomiality \cite{Ji:1998pc, Radyushkin:1998bz, Polyakov:1999gs}, they all general properties of GPDs. Furthermore, they have been also shown to account for partial conservation of the axial current \cite{Polyakov:1998ze, Mezrag:2014jka}, as one might expect in the case of the pion. Moreover, the resulting model exhibits agreement with available extractions for electromagnetic \cite{Huber:2008id, Amendolia:1986wj} and Gravitational Form Factors (GFFs) \cite{Kumano:2017lhr} of the pion, specially in the low-$t_{\pi}$ domain. Even more, the GPD thus obtained showed to be continuous along the $x=\xi$ line, as required by analyticity, for the amplitudes of DVCS to be finite \cite{Collins:2018aqt, EvolutionPaper}.

The resulting model puts the cap on a long effort developed during the last decade \cite{Mezrag:2013mya, Mezrag:2014tva, Mezrag:2014jka, Mezrag:2016hnp, Chouika:2017dhe, Chouika:2017rzs}, showing that a realistic description of pion's $3$D structure can be achieved from continuum Dyson-Schwinger computations. The model presented in \cite{Chavez:2021llq} exhibiting not only the formal requirements needed on a GPD, but also those of a phenomenological origin, constitutes a serious candidate for a realistic description of pion's structure. Its foundations on very first principles of quantum field theory together with its agreement with existing data for observables like EFFs, GFFs (specially in the region to be probed at EICs) or PDFs, points toward the possibility of obtaining solid predictions on the phenomenology to be observed at future colliders. Its use in the present study being self-justified and confidence on the model's independence of the results pushed forward.

\subsection{GPD evolution}

Given a formal description of pion's structure through GPDs, facing an analysis of the phenomenology to be observed at the foreseen EIC crucially requires QCD's scale-evolution to handled. The EIC will operate at energies in the range of tens of $\text{GeV}^{2}$s \cite{AbdulKhalek:2021gbh}, therefore running the GPD model from its definition scale to an experimentally relevant one is mandatory.

As described by \cite{Raya:2021zrz, Chavez:2021llq}, in the sense of scale-dependence, the model was built under a sole assumption: the existence of a scale, $\mu_{H}$, at which the entire parton content of the pion is described by two dressed valence-quarks. Such hypothesis has proved to yield successful descriptions at the PDF level \cite{Ding:2019lwe, Cui:2020tdf}. Moreover, in the case of the pion, it has been recently found that its existence is a solid prediction of available Lattice-QCD and Drell-Yan data \cite{Cui:2021mom, Raya:2021zrz}.

For these reasons we choose a description of pion's structure to be given at a reference scale by the $u$- and $\bar{d}$-GPDs developed in \cite{Chavez:2021llq}. We take GPD scale-evolution to be driven by the process independent effective coupling introduced in \cite{Rodriguez-Quintero:2018wma}, which accounts for saturation in the infrarred regime  by means of dynamical generation for a gluon mass-scale. Thus, the reference scale is set accordingly to $\mu_{H}=0.331~\text{GeV}$ \cite{Cui:2020tdf}, and the Leading Order (LO) QCD evolutions for GPDs solved through \texttt{Apfel++} software \cite{Bertone:2013vaa, Bertone:2017gds}.

\begin{figure}[H]
\includegraphics[scale=0.625]{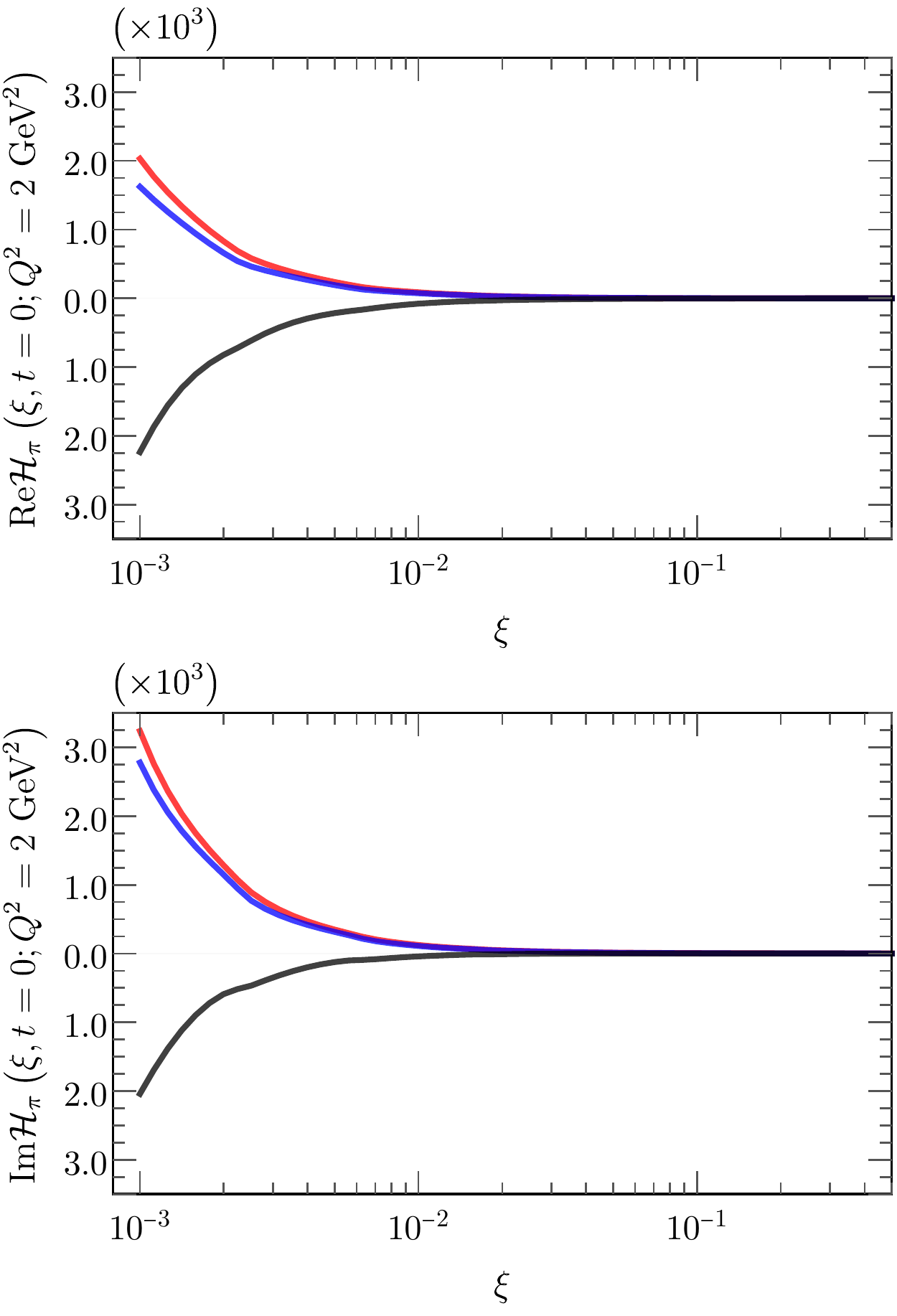}
\caption{\label{fig:CFFs} DVCS Compton Form Factors at a factorization scale of $2~\text{GeV}^{2}$. \textit{Upper panel} - Real part. \textit{Lower panel } - Imaginary part. \textit{Legend} - Red line: LO evaluation; blue line: NLO without gluon GPDs. Black line: full NLO result.}
\end{figure}

\section{Compton Form Factors}\label{sec:CFF}

With a realistic model for pion GPDs evolved to experimentally relevant energy scales, we are now in a position to employ them in computing CFFs, the coefficient functions parametrizing the amplitudes for DVCS \cite{Ji:1996nm, Kumericki:2016ehc, Pire:2011st, Moutarde:2013qs}:
\begin{equation}\label{eq:CFF}
\mathcal{H}_{\pi}\left(\xi,t,Q^{2}\right)=\hspace{-0.3cm}\sum_{i=\left\lbrace q\right\rbrace,g}\hspace{-0.25cm}\int_{-1}^{1}\frac{dx}{\xi}\mathcal{K}^{i}\hspace{-0.1cm}\left(\frac{x}{\xi},\frac{Q^{2}}{\mu_{F}^{2}}\right)H^{i}_{\pi}\left(x,\xi,t_{\pi};\mu_{F}^{2}\right)
\end{equation}
with the sum extended over all possible quark flavours and gluons; and where $\mathcal{K}^{i}$ is a convolution kernel calculable in perturbation theory.
\end{multicols}

\begin{figure}[H]
\centering
\includegraphics[scale=0.79]{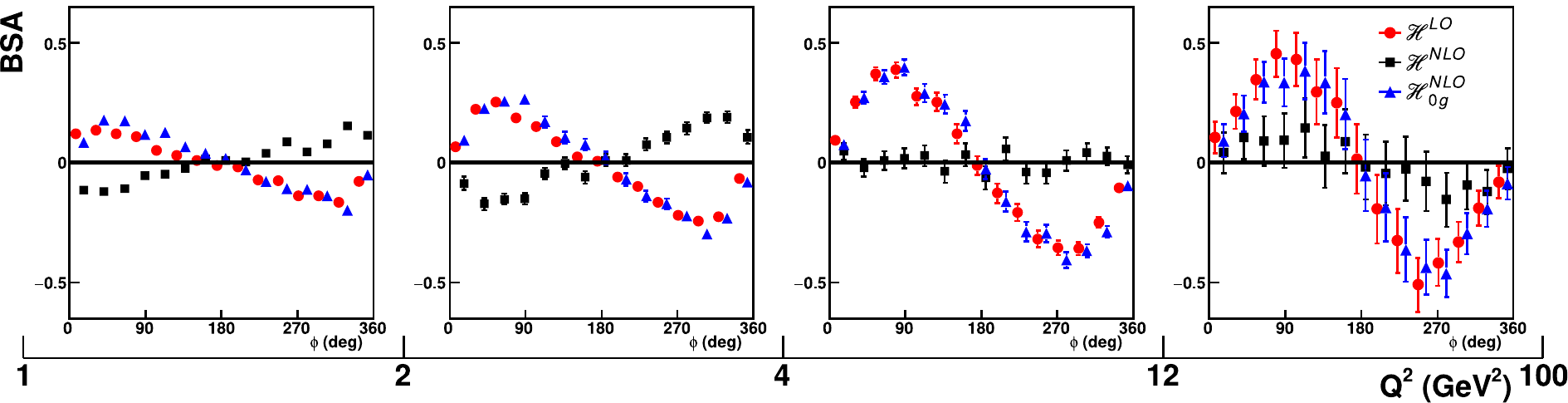}
\caption{\label{fig:Asymm} Expected lepton-beam-spin asymmetries as a function of $Q^{2}$ for $\xi\in\left[10^{-3},10^{-2}\right]$. \textit{Legend} - Red circles: LO evaluation; blue triangles: NLO without the gluon GPDs. Black squares: full NLO result. (Figure data from \cite{Chavez:2021koz}).}
\end{figure}

\begin{multicols}{2}
Interestingly, the gluon-type contribution to the convolution kernel, $\mathcal{K}^{g}\left(z\right)$, vanishes at LO. It is only from NLO on contributions in the strong running coupling when gluon content within the probed hadron enters the game \cite{Ji:1997nk, Pire:2011st, Moutarde:2013qs, Belitsky:1999sg}, affecting the behaviour of the hadrons subjected to deeply virtual Compton scattering. For this reason we compute  the pion's DVCS CFFs through three different approaches: (i) at LO, (ii) at NLO and finally, (iii) at NLO but manually setting the gluon contribution to zero.

With the help of \texttt{PARTONS} framework \cite{Berthou:2015oaw}, calculation of CFFs through these three approaches is feasible. As an illustration, results at a factorization scale of $2~\text{GeV}^{2}$ are shown in Fig. \ref{fig:CFFs}. There, it is clearly seen that both LO and NLO without gluon calculations, yield very similar results. The real and imaginary parts of the CFFs being positive definite and decreasing functions of the skewness. On the contrary, the full NLO result shows that gluon content within the pion gives the dominant contribution to the DVCS CFF. Specially in the low-$\xi$ region, which is the one to be probed at the EIC \cite{AbdulKhalek:2021gbh}. Crucially, the observed behaviour drastically changes. Even flipping sign for both, real and imaginary parts. In this respect, it is important to notice that gluon dominance yield a CFF roughly behaving as $\xi^{-1.4}$. A result which remains compatible with those independently obtained from DVCS dispersion relations with one subtraction constant \cite{Diehl:2007jb}; once again supporting, as argued in Sec. \ref{sec:GPDmodel}, supporting a limited model dependence of our predictions.

\section{Beam-spin asymmetries at the EIC}

Finally, after the development of the preceding sections, we are in a position to evaluate the cross-section for the Sullivan process. To this end we build a Monte Carlo event generation algorithm. We account for the prescriptions given in \cite{Amrath:2008vx}, in charge of assuring the interpretation of the process in the one-pion-exchange approximation; and also the EIC's detector characteristics specified in \cite{AbdulKhalek:2021gbh}. In that way, the phase-space of kinematic configurations is generated, and the formulae given in \cite{Amrath:2008vx} exploited for the corresponding evaluation of the cross-sections\footnote{For further details see \cite{Chavez:2021koz}.}. Finally, lepton's BSA are computed as defined in Eq. \eqref{eq:LBSA}.

The obtained results are shown in Fig. \ref{fig:Asymm} for $\xi\in\left[10^{-3},10^{-2}\right]$ and four different bins in $Q^{2}$. In agreement with those results discussed in Sec. \ref{sec:CFF}, LO and NLO without gluon contribution asymmetries are nearly indistinguishable. As expected, a sinusoidal shape is observed for the asymmetry at all explored scales \cite{Belitsky:2005qn}. Only an enhancement in their amplitudes is observed as the virtuality of the probing photon increases. In contrast, the result yielded by the full NLO calculation exhibits a markedly different behaviour. At low $Q^{2}$, as a manifestation of gluon-dominance, the asymmetry changes sign with respect to the LO and NLO/NoGluon ones. As photon's virtuality increases, its amplitude is now reduced; turning compatible with zero at moderate scales. Moreover, at high enough $Q^{2}$, the NLO result approaches the tendency of LO results, as expected from perturbation theory.

\vspace*{-0.15cm}

\section{Conclusions}

In the light of the results discussed herein, two main conclusions can be drawn. First, gluon content within the pion gives the dominant contribution to its behaviour in DVCS at low virtuality and small $\xi$. Second, a clearly non-vanishing lepton-beam-spin asymmetry is to be observed at the future EIC, thus opening the possibility to experimentally access, for the first time, pion's three dimensional structure; and even allowing to pin down the regime of gluon-dominance through an observable sign change.

\vspace*{-0.15cm}

\section*{Acknowledgements}
We would like to thank P. Barry, H. Dutrieux, T. Meisgny, B. Pire, K. Raya, C.D. Roberts, D. Sokhan, Q.-T. Song, P. Sznajder and J. Wagner for interesting discussions and stimulating comments. J.M.M.C acknowledges support from University of Huelva (grant EPIT-2021). F.S., J.R.Q. and J.S.'s work is supported by Ministerio de Ciencia e Innovaci\'on (Spain) under grant PID2019-107844GB-C22; Junta de Andaluc\'ia, under contract No. operativo FEDER Andaluc\'ia 2014-2020UHU-1264517 and P18-FR-5057 and PAIDI FQM-370. This work was also supported by the European Union's Horizon 2020 research and innovation programme under grant agreement No 824093; and through the GLUODYNAMICS project funded by "P2IO LabEx (ANR-10-LABX-0038)" in the framework "Investissements d'Avenir" managed by the Agence Nationale de la Recherche (ANR), France.

\end{multicols}
\medline
\begin{multicols}{2}
\bibliography{Bibliography}
\bibliographystyle{unsrt}
\end{multicols}
\end{document}